\documentclass[twocolumn,english]{revtex4-1}
\usepackage[T1]{fontenc}
\usepackage[latin9]{inputenc}
\usepackage{babel}
\usepackage{float}
\usepackage{amsmath}
\usepackage{graphicx}
\usepackage[unicode=true,
 bookmarks=true,bookmarksnumbered=false,bookmarksopen=false,
 breaklinks=false,pdfborder={0 0 1},backref=false,colorlinks=false]
 {hyperref}

\makeatletter
\usepackage{hyperref}

\makeatother

\begin{document}
\title{Scaling behavior of transient dynamics of vortex-like states in self-propelled
particles}
\author{Pei-fang Wu}
\thanks{These two authors contributed equally to this work.}
\affiliation{Guangdong Provincial Key Laboratory of Quantum Engineering and Quantum
Materials, School of Physics and Telecommunication Engineering, South
China Normal University, Guangzhou 510006, China.}
\author{Wei-chen Guo}
\thanks{These two authors contributed equally to this work.}
\affiliation{Guangdong Provincial Key Laboratory of Quantum Engineering and Quantum
Materials, School of Physics and Telecommunication Engineering, South
China Normal University, Guangzhou 510006, China.}
\author{Bao-quan Ai}
\email{aibq@scnu.edu.cn}

\affiliation{Guangdong Provincial Key Laboratory of Quantum Engineering and Quantum
Materials, School of Physics and Telecommunication Engineering, South
China Normal University, Guangzhou 510006, China.}
\author{Liang He}
\email{liang.he@scnu.edu.cn}

\affiliation{Guangdong Provincial Key Laboratory of Quantum Engineering and Quantum
Materials, School of Physics and Telecommunication Engineering, South
China Normal University, Guangzhou 510006, China.}
\begin{abstract}
\noindent \textbf{Abstract}: Nonequilibrium many-body transient dynamics
play an important role in the adaptation of active matter systems
to environmental changes. However, the generic universal behavior
of such dynamics is usually elusive and left as open questions. Here,
we investigate the transient dynamics of vortex-like states in a two-dimensional
active matter system that consists of self-propelled particles with
alignment interactions subjected to extrinsic environmental noise.
We identify a universal power-law scaling for the average lifetime
of vortex-like states with respect to the speed of the self-propelled
particles. This universal scaling behavior manifests strong robustness
against the noise, up to the level where influences from environmental
fluctuations are large enough to directly randomize the moving directions
of particles. Direct experimental observations can be readily performed
by related experimental setups operated at a decently low noise level.
\end{abstract}
\maketitle

\section{Introduction}

Active matter composed of many self-propelled agents is a class of
intrinsic nonequilibrium (NEQ) systems whose length scale spans from
oceanic to subcellular, while the agents can be generally modeled
as self-propelled particles \citep{Popkin_Nature_2016,Ramaswamy_ARCMP_2010,Marchetti_RMP_2013,Bechinger_RMP_2016,Chate_ARCMP_2020,Vicsek_Phys_Rep_2012}.
In these NEQ systems, various collective behavior has been identified
\citep{Cavagna_ARCMP_2014,Cavagna_Phys_Rep_2018,Pearce_PNAS_2014,Ginelli_PNAS_2015,Bottinelli_PRL_2016,Bain_Science_2019,Schaller_Nature_2010,Dombrowski_PRL_2004,Zhang_PNAS_2010,Buttinoni_PRL_2013,Palacci_Science_2013,Kursten_PRL_2020_02,Kursten_PRL_2020_10},
and the transient dynamics of changing from one collective behavior
to another, either spontaneously or in reaction to external perturbations,
are also observed frequently \citep{Parrish_Science_1999,Couzin_ASB_2003,Becco_Physica_A_2006}.
Such transient dynamics play an important role in the adaptation
of active matter systems to environmental changes. But in most cases,
the universal scaling behavior of these transient dynamics is rather
elusive and still left as open questions.

A case in point is associated with the vortices or vortex-like states,
which frequently appear and undergo transient dynamics with a finite
lifetime, resulting in rich dynamical behavior not only in classical
\citep{Landau_Book_Fluid_Mechanics_1987} and quantum fluids \citep{Leggett_Book_2006},
but also in active matter systems \citep{Vicsek_Phys_Rep_2012,Duan_PRE_2015,Chen_PRE_2012,Cheng_NJP_2016,costanzo_JPD_2018},
ranging from animal flocks, over active colloids \citep{Kokot_Nat_Com_2018},
to bacterial colonies \citep{Wioland_Nat_Phys_2016,Chen_Nature_2017,Dunkel_PRL_2013,Grossmann_PRL_2014,nishiguchi_Nat_Com_2018,liu_Nat_2021}
and collectively moving microtubules \citep{Vicsek_Nature_2012,Sumino_Nature_2012},
etc. In addition, the vortices or vortex-like states have also been
observed in active matter systems without alignment interactions,
such as active particles in phase-separated configurations, homogeneous
dense systems, narrow circular crowns \citep{Caprini_PRL_2020,Caprini_L_PRL_2020,Caprini_Chem_2021},
etc. Despite generic universal behavior in these many-body transient
dynamics being generally quite elusive due to their intrinsic NEQ
characteristic, some can indeed be identified, for instance, the famous
$-5\slash3$ power-law scaling for the energy spectrum of turbulence
identified by Kolmogorov \citep{Kolmogorov_DANSSSR_1941,Frisch_Book_1995}.
Noticing also that certain universal dynamical scaling behavior could
emerge in NEQ many-body systems approaching their equilibrium counterparts
\citep{Sieberer_PRL_2013,Tauber_PRX_2014}, this thus raises the intriguing
question of whether certain universal scaling behavior exists in the
transient dynamics of vortex-like states in active matter systems
close to equilibrium.

In this work, we address this question for two-dimensional (2D) active
matter systems that consist of self-propelled particles with alignment
interactions subjected to extrinsic environmental noise {[}cf.~Eq.~(\ref{eq:Direction_with_extrinsic_noise}){]}.
In particular, we focus on the near-equilibrium regime of the system
where the speed $v$ of the self-propelled particles is low \citep{Toner_PRL_1995}.
By systematically investigating the transient dynamics of vortex-like
states in this regime, we identify a universal power-law scaling
of the average lifetime $t^{*}$ of vortex-like states with respect
to the speed $v$ of the self-propelled particles, i.e., $t^{*}\propto v^{-\alpha}$
with $\alpha\simeq1$ (cf.~Fig.~\ref{Fig_weak_noise}). This scaling
behavior originates from the occurrence of ``separated particles''
{[}cf.~Fig.~\ref{Fig_weak_noise}(c){]} and manifests strong robustness
against the environmental noise up to the level where environmental
fluctuations are large enough to directly randomize the moving directions
of particles (cf.~Fig.~\ref{Fig_strong_noise}). The upper bound
of this noise level is further estimated by calculating the survival
probability of the vortex-states in the $\upsilon\rightarrow0$ limit
(cf.~Fig.~\ref{Fig_critcal_noise_estimate}). In regard to the robustness
of the power-law scaling against environmental fluctuations, we expect
it can be readily observed in current experimental setups such as
in animal groups in quasi-2D space, e.g., fishes in a shallow tank,
by employing video tracking techniques, or in 2D synthetic active
matter systems consisting of self-propelled colloidal particles with
alignment interactions \citep{Bricard_Nat_Com_2015}.

The rest of the paper is organized as follows. In Sec.~\ref{sec:System-and-Method},
we specify the system and model under study. In Sec.~\ref{sec:Results},
we discuss the power-law scaling behavior of the transient dynamics
at low noise levels and the typical behavior of the transient dynamics
at high noise levels. In Sec.~\ref{sec:Experimental-observability},
we discuss possible experimental observations of the power-law scaling
behavior identified. Finally, we conclude and give an outlook in Sec.~\ref{sec:Conclusion}.

\section{System and Model\label{sec:System-and-Method}}

The system under study consists of $N$ self-propelled particles moving
in 2D with a constant speed $\upsilon$, whose dynamics is modeled
by the Vicsek model with extrinsic noise \citep{Gregoire_PRL_2004,Chate_PRE_2008,Caussin_PRL_2014,Solon_PRL_2015},
\begin{align}
\boldsymbol{x}_{j}(t+\Delta t) & =\boldsymbol{x}_{j}(t)+\boldsymbol{v}_{j}(t)\Delta t,\label{eq:Position}\\
\theta_{j}(t+\Delta t) & =\arg[\sum_{k\in U_{j}}(e^{i\theta_{k}(t)}+\eta e^{i\xi_{j}(t)})]\label{eq:Direction_with_extrinsic_noise}
\end{align}

Here, $\Delta t$ is a discrete time step, $\boldsymbol{x}_{j}(t)$,
$\boldsymbol{v}_{j}(t)$ and $\theta_{j}(t)$ are the position, the
velocity and the direction of motion of the $j$th particle at time
$t$, respectively. $\xi_{j}(t)\in[-\pi,\pi]$ is a uniformly distributed
random noise with $\eta\in[0,1]$ being the extrinsic noise level.
These particles interact with each other via the alignment interaction,
i.e., for instance, the $j$th particle tends to move along the average
direction of its neighbors $U_{j}$ within a circular region of radius
$r$. Due to the competition between the alignment interaction and
the environmental fluctuations, the steady state of this system assumes
an ordered flocking phase at low noise levels and a disordered phase
at high noise levels \citep{Gregoire_PRL_2004,Chate_PRE_2008}.

In fact, the above system also assumes a close relationship to the
2D XY model, with the velocity of the particle playing the role of
the local spin of the 2D XY model \citep{Toner_PRL_1995,Toner_Ann_Phys_2005,Kosterlitz_J_Phys_C_1973,Kosterlitz_Rep_Prog_Phys_2016}.
More specifically, in the low speed limit $\upsilon\rightarrow0$,
the dynamics of this NEQ system reduce precisely to Monte Carlo dynamics
of the 2D XY model. Noticing that at low temperatures (or equivalently,
low noise levels), the vortex-type configurations in the 2D XY model
are stable due to the topological protection \citep{Mermin_RMP_1979},
one would expect that in the low speed limit, vortex-like states of
the self-propelled particles in the 2D space could assume long-diverging
lifetime in their transient dynamics. This thus suggests the intriguing
possibility that certain universal scaling behavior could exist in
the transient dynamics of vortex-like states in the active matter
system under study. Indeed, as we shall see in the following, the
average lifetime $t^{*}$ of vortex-like states in this system manifests
a universal power-law scaling with respect to the speed $v$ at relatively
low noise levels.

\section{Results\label{sec:Results}}

In the following, we focus on how the average lifetime of the vortex-like
state changes with respect to the speed of the self-propelled particles
at different noise levels. To this end, we numerically simulate the
dynamics of the system with its initial states prepared in the vortex-like
states, and monitor their dynamics at different self-propelling speeds
$v$ and noise level $\eta$. More specifically, here we follow a
simple protocol where the system is first initialized with the type
of vortex-like states which can be relatively easily prepared in various
experimental setups \citep{Bricard_Nat_Com_2015}, where each of the
$N$ self-propelled particles is randomly located on a circle with
a radius $R$ with the direction of its velocity aligned with the
local tangent direction of the circle {[}cf.~the first plot from
the left of Fig.~1(a){]}. For each stochastic trajectory of the transient
dynamics of the vortex-like state, the lifetime $t_{n}^{*}$ for the
vortex-like state, with $n$ being the trajectory index, is extracted
by monitoring the instant winding number $w(t)=\sum_{j=0}^{N}(\theta_{j+1}(t)-\theta_{j}(t))\slash2\pi$
of the system, which equals to one for the initial state, i.e., $w(t=0)=1$,
and jumps to zero at the time point when the vortex-like state disappear,
i.e., $w(t=t_{n}^{*})=0$ {[}cf.~for instance the fourth plot from
the left of Fig.~1(a){]}. In particular, since the self-propelled
particles are randomly located on a circle, the $\left(N+1\right)$th
particle is equivalent to the $0$th particle. The average lifetime
$t^{*}$ of the vortex-like states is calculated by $t^{*}=\sum_{n=1}^{N_{{\rm traj}}}t_{n}^{*}\slash N_{{\rm traj}}$.
We remark that at the low noise level below the flocking transition
point, the final steady state of the system is a flocking state, i.e.,
all particles moving in the same direction as shown in the rightmost
snapshot in Fig.~1(a). Moreover, Fig.~1(a) only shows the time-evolution
that corresponds to one of the stochastic trajectories. For different
stochastic trajectories, the directions for the final collective motion
are different. If not specified in text, we use $N_{\mathrm{traj}}=1.5\times10^{2}$
stochastic trajectories to perform ensemble averages and set $r=1,R=5,\Delta t=1$.

\subsection{Power-law scaling behavior in the transient dynamics at low noise
levels\label{subsec:Low-noise-level}}

\begin{figure}[H]
\noindent \begin{centering}
\includegraphics[width=3.3in]{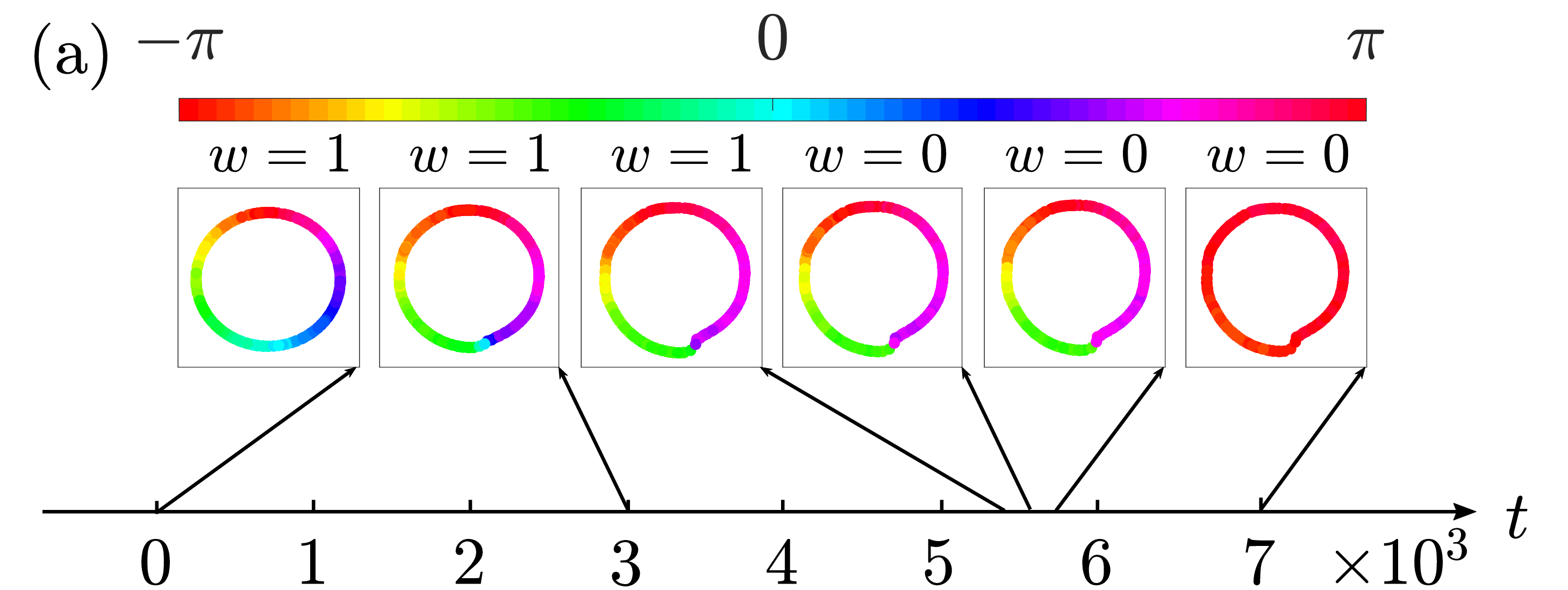}
\par\end{centering}
\noindent \begin{centering}
\includegraphics[width=3.3in]{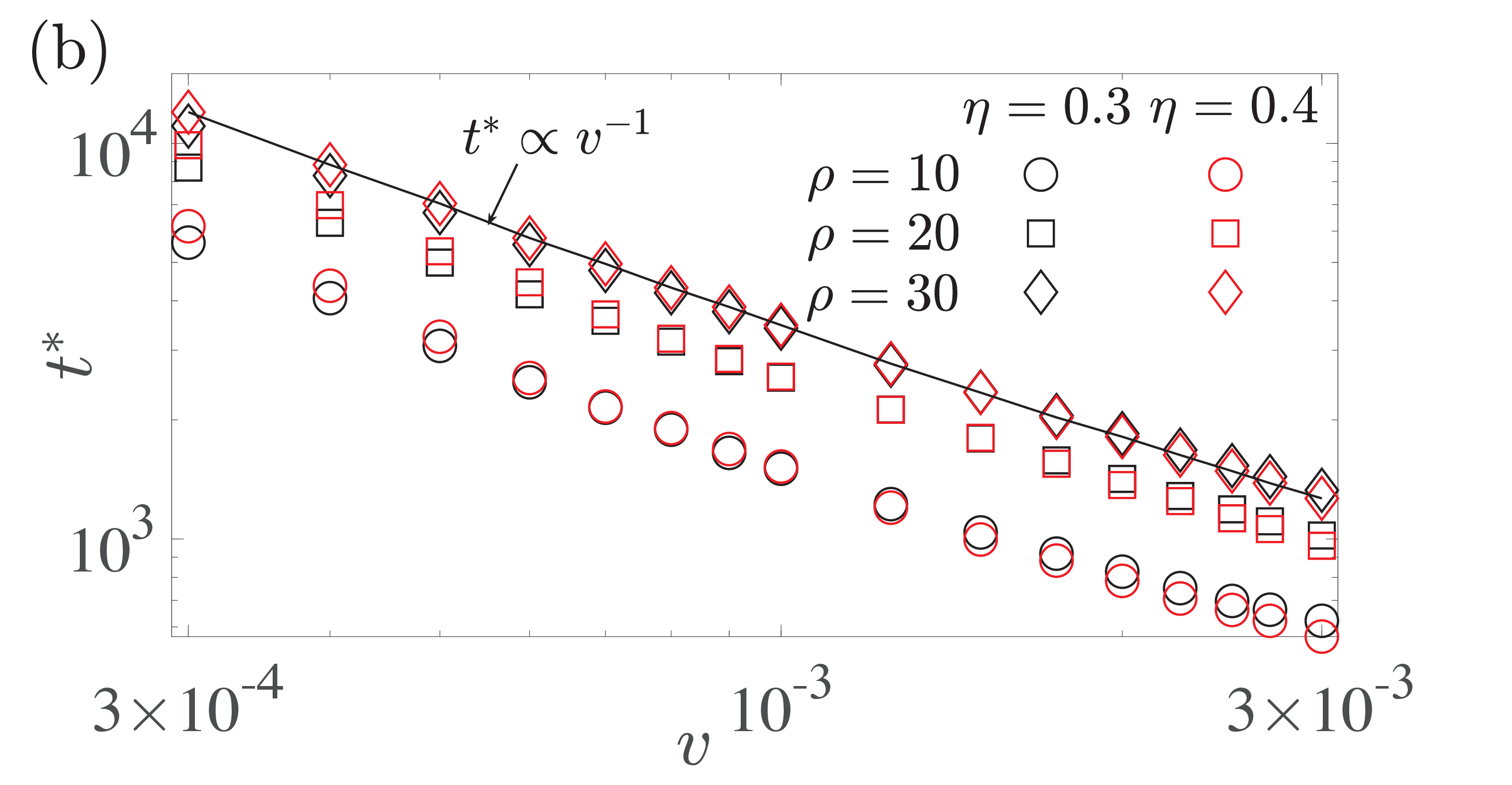}
\par\end{centering}
\noindent \begin{centering}
\includegraphics[width=3.3in]{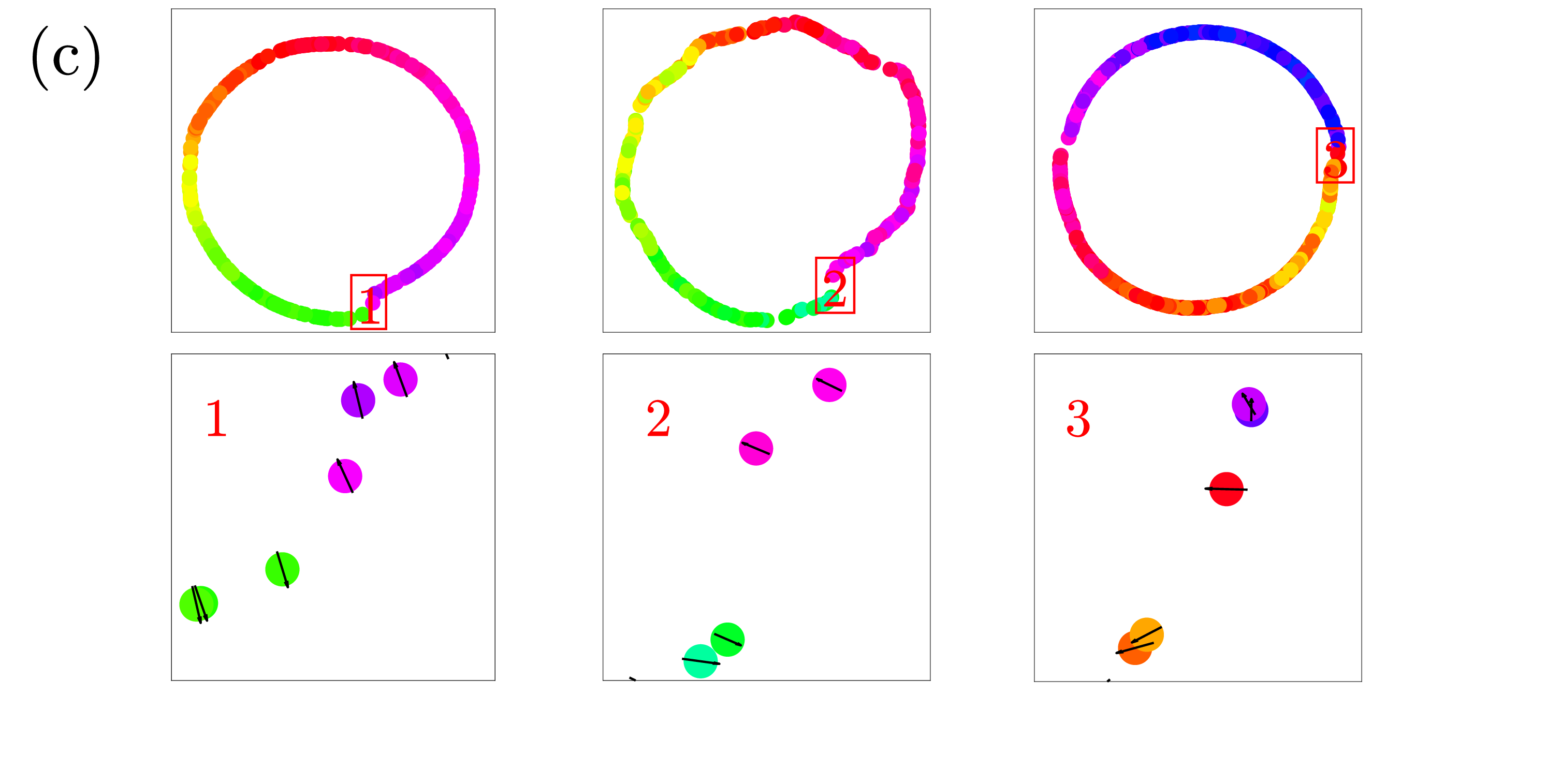}
\par\end{centering}
\caption{\label{Fig_weak_noise}(a) Typical transient dynamics of vortex-like
states in self-propelled particles at low noise levels. The position
of each circle represents the position of each self-propelled particle
in the coordinate space, and the color represents the direction of
motion of each particle. It gradually deforms and disappears after
some period of time. (b) Power-law scaling behavior of the transient
dynamics at low noise levels. The black (red) marks correspond to
the noise level $\eta=0.3$ ($\eta=0.4$). The average lifetime $t^{*}$
of vortex-like states manifests a power-law dependence on $v$ in
the low speed limit, i.e., $t^{*}\propto v^{-\alpha}$ with $\alpha\simeq1$.
More specifically, values of the scaling exponent $\alpha$ extracted
from the power-law fitting of the data are $\alpha=1.06,\,1.09,\,0.99,\,1.09,\,0.99,\,1.05$
for $(\rho,\eta)=(10,0.3),\,(10,0.4),\,(20,0.3),\,(20,0.4),\,(30,0.4)$,
respectively. (c) Typical configurations right before vortex-like
states disappear at low noise levels. The upper figures show the complete
configurations and the lower ones show the corresponding zoom-in of
the indexed rectangle area. The arrow also represents the direction
of motion of each particle. These configurations indicate that the
origin of the scaling behavior is associated with the occurrence of
separated particles that are more inclined to be influenced by environmental
fluctuations. See text for more details.}
\end{figure}

Fig.~\ref{Fig_weak_noise}(b) shows the speed $v$ dependence of
the average lifetime $t^{*}$ at different system parameters in the
low noise level regime. Here, the average lifetime $t^{*}$ of vortex-like
states manifests a power-law dependence of $v$ , i.e., $t\propto v^{-\alpha}$,
in the low speed limit in double logarithmic coordinates. We further
extracted the scaling exponent $\alpha$ from the power-law fitting
of the data and find their value are around $\alpha=1$ {[}$\alpha=1.06$
for $(\rho=10,\eta=0.3)$, $\alpha=1.09$ for $(\rho=10,\eta=0.4)$,
$\alpha=0.99$ for $(\rho=20,\eta=0.3)$, $\alpha=1.09$ for $(\rho=20,\eta=0.4)$,
$\alpha=0.99$ for $(\rho=30,\eta=0.3)$, $\alpha=1.05$ for $(\rho=30,\eta=0.4)$,
and $\rho$ is the initial particle number density on arc, i.e., $\rho=N/(2\pi R)${]}.
This suggests the existence of a universal power scaling $t^{*}\propto v^{-1}$.

To understand the origin of this scaling behavior, we monitor the
evolution of the system configuration in the transient dynamics. We
find the vortex-like state disappears after ``separated'' particles
(particles that only have few neighbors) occur {[}cf.~Fig.~\ref{Fig_weak_noise}(c)
for typical configurations right before the vortex-like state disappears{]}.
Indeed, it is natural to expect that the \textquotedblleft separated\textquotedblright{}
particles are more inclined to be influenced by environment fluctuations
since the alignment interaction among them is very weak due to the
lack of neighbors, hence they are expected to play a key role in causing
the disappearance of vortex-like states. Since a portion of self-propelled
particles need to travel a certain distance $l$ to separate from
each other, this separation process is naturally expected to take
a period of $l/v$, after which the vortex-like states disappear.
This thus suggests that the average lifetime $t^{*}$ of the vortex-like
states should be proportional to $\upsilon^{-1}$ which indeed matches
the scaling observed in numerical simulations.

Moreover, we also notice that this scaling is robust (the scaling
exponent $\alpha$ deviates from $1$ by no more than $5\%$) at even
moderate noise levels ($1/2<\eta/\eta_{c}<1$ with $\eta_{c}\sim0.6$
being the typical critical noise level of the flocking transition)
{[}cf.~the red marks in Fig.~\ref{Fig_weak_noise}(b){]}. This is
consistent with the expectation from the close relationship between
the system and the 2D XY model in the low speed limit ($v/(r/\Delta t)\ll1$),
where vortex-type configurations in the 2D XY model are robust against
thermal fluctuations due to the topological protection \citep{Mermin_RMP_1979}.
Noticing that at large enough noise level or high enough temperature
($T/T_{c}>1$ with $T_{c}$ being the critical temperature of the
2D XY model) the 2D XY model assumes a disordered phase \citep{Kosterlitz_J_Phys_C_1973,Kosterlitz_Rep_Prog_Phys_2016}
with free vortices continuously generated and annihilated in its Monte
Carlo dynamics, and from the close relationship between the system
and the 2D XY model in the low speed limit \citep{Toner_PRL_1995,Toner_Ann_Phys_2005},
we therefore expect that the power-law scaling of the system's transient
dynamics at moderate noise levels should be absent at high enough
noise levels ($\eta/\eta_{c}>1$), as we shall now discuss in the
following.

\begin{figure}
\noindent \begin{centering}
\includegraphics[width=3.3in]{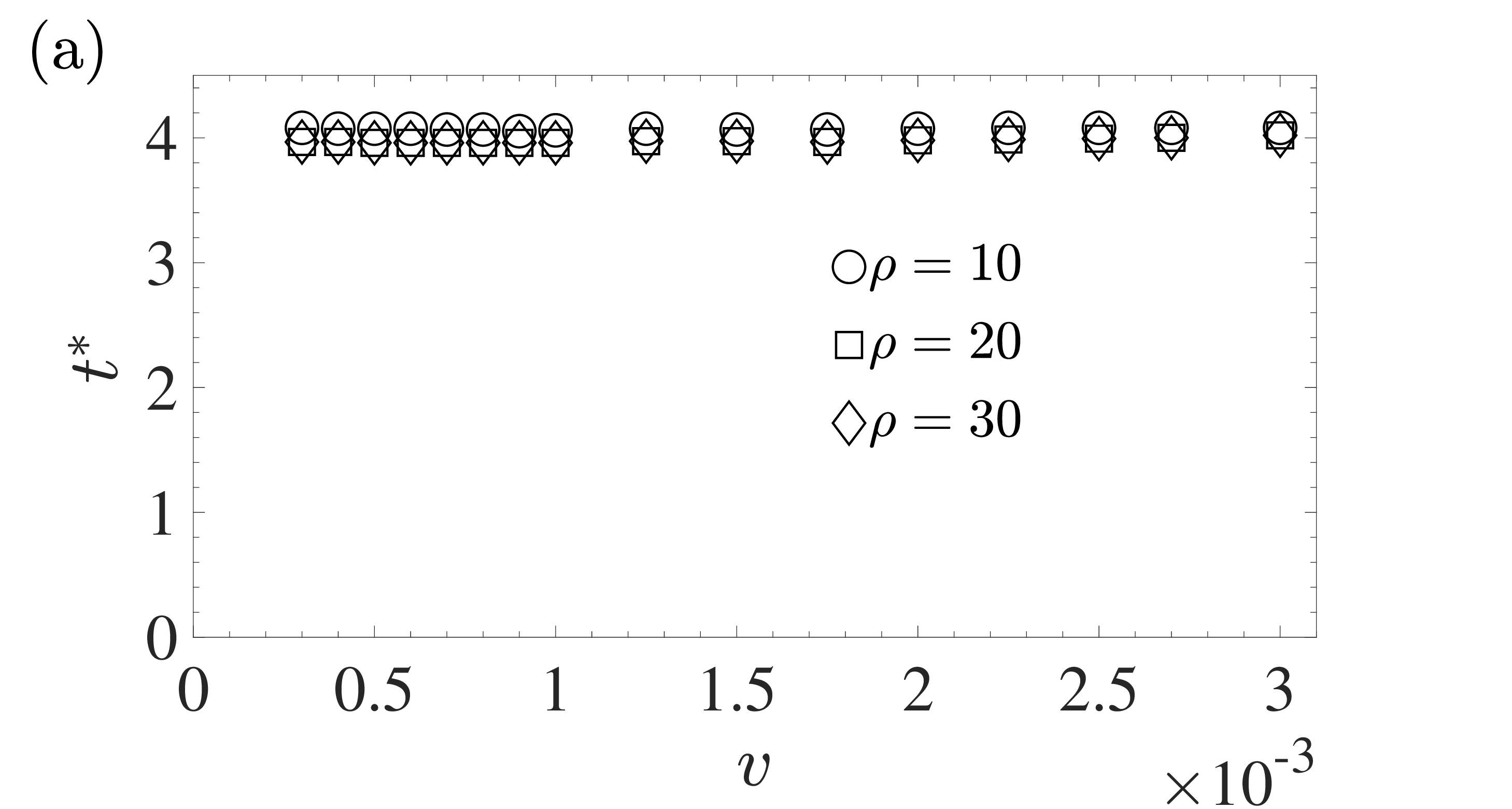}
\par\end{centering}
\noindent \begin{centering}
\includegraphics[width=3.3in]{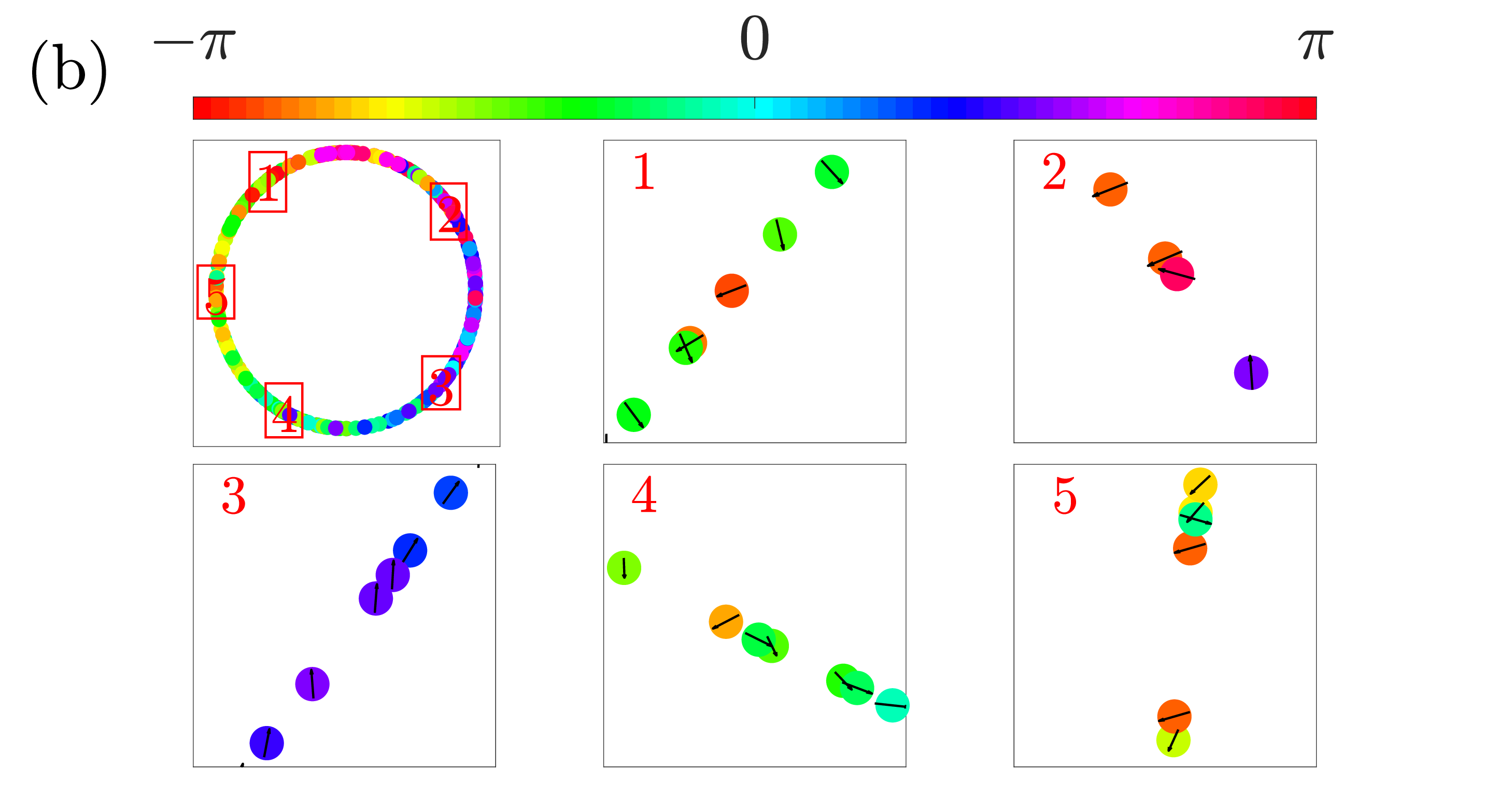}
\par\end{centering}
\caption{\label{Fig_strong_noise}(a) Absence of the power-law scaling in transient
dynamics at a high noise level with $\eta=0.7$. In the presence of
strong environmental fluctuations, vortex-like states quickly disappear
and the average lifetime $t^{*}$ essentially does not depend on the
speed $\upsilon$ of self-propelled particles. (b) Typical configuration
right before the vortex-like state disappears at a high noise level
with $\eta=0.7$. The first plot shows a typical configuration and
other plots correspond to the zoom-in of the indexed rectangle area,
respectively. In this case, environmental fluctuations are large enough
to directly randomize the moving direction of each particle in a short
time. See text for more details.}
\end{figure}

\begin{figure}
\noindent \begin{centering}
\includegraphics[width=3.3in]{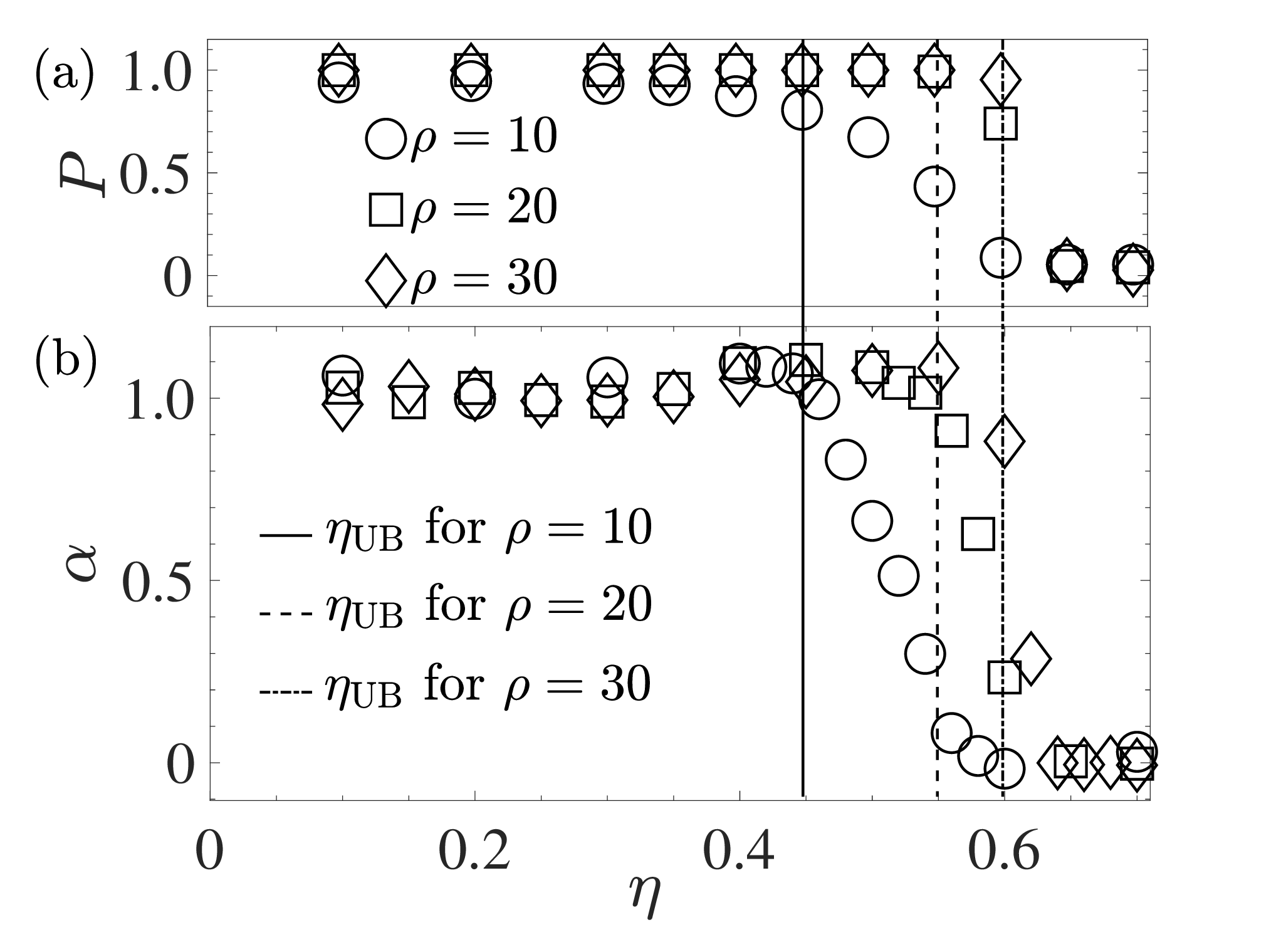}
\par\end{centering}
\caption{\label{Fig_critcal_noise_estimate}(a) Survival probability $P$ of
the vortex-states as a function of the noise level at zero speed $v=0$.
For each set of system parameters, $N_{\mathrm{traj}}=10^{5}$ different
initial configurations of vortex-like states are evolved for a fixed
period of time $T=10^{3}$, respectively. The survival probability
$P$ undergoes a fast decay from $1$ to $0$ once the noise level
$\eta$ exceeds a critical value $\eta_{\mathrm{UB}}$ (vertical black
lines), indicating that for $\eta>\eta_{\mathrm{UB}}$ the influences
of environmental fluctuations are so large that the power scaling
behavior $t^{*}\propto v^{-1}$ does not exist anymore. (b) Extracted
values for the exponent $\alpha$ from the power law fit the $\upsilon$
dependence of $t^{*}$ at different noise levels $\eta$. For $\eta>\eta_{\mathrm{UB}}$,
$\alpha$ manifests large deviation from the value $\alpha=1$. See
text for more details.}
\end{figure}

\subsection{Transient dynamics at high noise levels\label{subsec:high_noise_level}}

Fig.~\ref{Fig_strong_noise}(a) shows the speed $v$ dependence of
the average lifetime $t^{*}$ in the high noise level regime with
$\eta=0.7$. Here, one can notice that the vortex-like states quickly
disappear and the average lifetime $t^{*}$ essentially does not depend
on the speed $\upsilon$ of self-propelled particles, indicating that
the corresponding power scaling behavior is absent in the presence
of strong environmental fluctuations. To further investigate the dynamical
behavior of the vortex-like states in the high noise level regime,
we monitor the evolution of the system configurations. As we can see
from Fig.~\ref{Fig_strong_noise}(b), which shows a typical configuration
of the system at the moment right before the disappearance of the
vortex-like state, the moving directions of most self-propelled particles
are strongly influenced by the strong environmental fluctuations and
become disordered in a short time, resulting in the disappearance
of the vortex-like state. This is in sharp contrast to the typical
dynamical behavior in the low noise regime where the disappearance
of the vortex is accompanied by spatial deformations {[}cf.~Fig.~\ref{Fig_weak_noise}(c){]}.

So far we have seen the transient dynamics of vortex-states show distinct
behavior at high and low noise regimes. To further quantitatively
distinguish the high noise level regime from the low noise level one,
we estimate the upper bound of the noise level above which environmental
fluctuations can easily make moving directions of most self-propelled
particles disordered, hence resulting in the absence of the scaling
observed in the low noise regime. We estimate this upper bound for
each set of fixed system parameters by investigating the survival
probability $P\equiv N_{w}/N_{\mathrm{traj}}$ of the vortex-states
with zero speed, i.e., $v=0$, in the presence of environmental noise.
Here, $N_{\mathrm{traj}}$ is the total number of trajectories {[}usually
$\sim\mathcal{O}(10^{5})${]} that are initialized with different
vortex-like states, and evolve for a fixed long period of time $T$
{[}usually $\sim\mathcal{O}(10^{3})${]}. $N_{w}$ is the number of
trajectories with $w(T)=1$, i.e. still remaining in the vortex-like
states at $t=T$. As one can see from Fig.~\ref{Fig_critcal_noise_estimate}(a)
that shows the dependence of $P$ on the noise level $\eta$ at three
sets of system parameters, the survival probability $P$ undergoes
a fast decay once the noise level $\eta$ exceeds a critical value
$\eta_{\mathrm{UB}}$, indicating that for $\eta>\eta_{\mathrm{UB}}$
the influences of environmental fluctuations are so large that the
power scaling behavior $t^{*}\propto v^{-1}$ does not exist anymore.
Indeed, as one can see from Fig.~\ref{Fig_critcal_noise_estimate}(b)
which shows the extracted exponents from the power law fit the $\upsilon$
dependence of $t^{*}$ at different noise levels $\eta$, the extracted
exponents manifest large deviation from the value $\alpha=1$ for
the noise levels above $\eta_{\mathrm{UB}}$, clearly showing that
for $\eta>\eta_{\mathrm{UB}}$ the power scaling behavior $t^{*}\propto v^{-1}$
does not exist anymore.

\section{Experimental observability\label{sec:Experimental-observability}}

We expect the predicted power-law scaling for the average lifetime
of the vortex-like states with respect to the speed $v$ of self-propelled
particles can be observed in current experimental setups. For instance,
one can employ the experimental setup presented in Ref.~\citep{Bricard_Nat_Com_2015},
where polymethyl methacrylate spheres are dispersed in hexadecane
solution. In the case where the colloid packing fraction is much smaller
than the area fraction, the physics of the system is captured by the
dynamical model investigated in this work. In the restricted space,
the self-propelled colloidal particles can form stable vortex-like
state configurations at the boundary \citep{Bricard_Nat_Com_2015},
and their transient dynamics can thus be investigated after removing
the boundary (actually, removing the boundary of the whole system
may not be easy to achieve directly in the experimental setup presented
in Ref.~\citep{Bricard_Nat_Com_2015}, however, one could further
engineer an additional removable boundary, for instance, a removable
ring-shaped barrier in the central region of the experimental setup
presented in Ref.~\citep{Bricard_Nat_Com_2015}). Here, the speed
of self-propelled particles $\upsilon\propto\sqrt{E^{2}\slash E_{c}^{2}-1}$
in experiments \citep{Lu_Soft_Matter_2018,Bricard_Nat_Com_2015},
with $E_{c}$ being a critical electrical field and $E$ being electric
field generated by the longitudinal voltage. Therefore, for experiments
operated in the low temperature regime (corresponding to low noise
levels), we expect that the power-law scaling can be observed experimentally
by monitoring the transient dynamics at different speeds of the self-propelled
particles tuned by the longitudinal voltage. Moreover, one can also
employ systems of bacteria. For instance, in the experiments presented
in Ref.~\citep{nishiguchi_Nat_Com_2018}, the vortex-like states
can form in the region surrounded by the neighboring pillars. Therefore,
if the pillars are further engineered to be removable in this experimental
setup, transient dynamics of the vortex-like states can be investigated
by removing the pillars. 

\section{Conclusions\label{sec:Conclusion}}

Despite the generic universal behavior of many-body transient dynamics
are generally quite elusive due to their intrinsic NEQ characteristic,
some can indeed be identified as the transient dynamics of vortex-like
states in self-propelled particles shows: The average lifetime of
the vortex-like states in this 2D active matter system manifests a
power-law dependence on the speed of the self-propelled particles
in a wide system parameter regime. In particular, this scaling behavior
is robust against the environmental fluctuations up to the finite
noise level where moving directions of self-propelled particles can
be directly randomized by the noise, indicating it is promising to
be observed directly in related experiments. Moreover, the environmental
noise employed in this work is not spatially correlated. In the case
where the spatial correlation length of the noise is relatively small
compared with the average separation between particles, the dynamical
behavior of the system is expected to show no substantial difference
from the one reported here. However, since in general, the environmental
noise can show some spatial correlations on a length scale larger
than the average separation between particles, it is quite intriguing
to investigate the transient dynamics of the system in this case.
We believe that our work will stimulate further theoretical and experimental
efforts in revealing the generic universal behavior of transient dynamics
in active matter systems. 
\begin{acknowledgments}
This work was supported by NSFC (Grant Nos.~11874017, 12075090, and
12275089), NKRDPC (Grant~No.~2022YFA1405304), GDSTC (Grant No.~2018A030313853
and No.~2017A030313029), GDUPS (2016), Major Basic Research Project
of Guangdong Province (Grant No.~2017KZDXM024), and START grant of
South China Normal University.
\end{acknowledgments}

\end{document}